\documentclass[letterpaper,twocolumn,10pt]{article}
\usepackage{usenix}

\usepackage[utf8]{inputenc}
\usepackage[english]{babel}
\usepackage{blindtext}
\usepackage{textcomp}
\usepackage{amsmath,amssymb,amsfonts}
\usepackage{makecell}
\usepackage{graphicx}
\usepackage{comment}
\usepackage{multirow}
\usepackage{fancyhdr}
\usepackage{xspace}
\usepackage{xcolor}
\usepackage{dsfont}
\usepackage{pgfplots}
\usepackage{multirow}
\usepackage{float}
\usepackage{microtype}
\usepackage{subfig}
\usepackage{booktabs}
\usepackage{hyperref}
\usepackage{algorithmic}
\usepackage{bussproofs}
\usepackage{mathtools}

\mathtoolsset{centercolon}

\pgfplotsset{compat=1.17}

\def \OurApproach{Ada-Grouper\xspace}
\def \PreviousWork{Rhino\xspace}
\def \OurSchedule{$k$F$k$B\xspace}

\title{\OurApproach: Accelerating Pipeline Parallelism in Preempted Network by Adaptive Group-Scheduling for Micro-Batches}
\author{
{\rm Siyu Wang} \qquad
{\rm Zongyan Cao} \qquad
{\rm Chang Si} \qquad
{\rm Lansong Diao} \qquad
{\rm Jiamang Wang} \qquad
{\rm Wei Lin} \\ \\
Alibaba Group \qquad
}
\begin{document}

\maketitle

\begin{abstract}
Pipeline parallelism has been demonstrated to be a remarkable approach to improve throughput for training deep neural networks with billions of parameters over heterogeneous clusters. The 1F1B scheduling plan is a widely adopted strategy for memory and performance optimization, which interchanges the forward and backward stage computations of different micro-batches. On the other hand, a common issue in using the 1F1B scheduling is that stage computation is delayed due to the data transfer when network resources are preempted by other tasks, even with the minimum communication between stages. The exclusive access of these network resources cannot be guaranteed in cloud offerings. We present a general scheduling technique to accommodate pipeline parallelism to preempted network environments at the expense of a certain amount of memory pressure. The core concept is to extend 1F1B schedule scheme to \OurSchedule, which groups $k$ micro-batches, and alternately executes $k$ forward and backward computations. We propose \OurApproach, an adaptive \OurSchedule scheduler which regularly adjusts the number of group members $k$ to maintain an optimal balance between communication and computation efficiency correspond to changes in a changing network environment under the memory limit. Experimental results demonstrate that our design maintains stable performance for pipeline parallelism, yielding a performance increase of up from 4\% to 30\%, compared with 1F1B in preempted network scenarios.
\end{abstract}

\section{Introduction}

Deep neural networks(DNNs) has expanded quickly in number of parameters. For natural language processing, GPT-3\cite{gpt3} has more than 175 billion parameters and PaLM\cite{palm} consists of 540B parameters. For multimodal pretraining, M6\cite{m6} has over 100 billion parameters. Recent studies have proved that large models exhibited unprecedented performance on extremely complex tasks, stimulating an even greater demand for larger model sizes. To train these large models effectively, distribution strategy is necessary. This involves partitioning the models or training data into different devices and executing them in parallel.

Several distribution strategies have been developed for training DNNs, such as pipeline parallelism\cite{gpipe}, which is particularly beneficial for training on heterogeneous clusters with weaker inter-connections(e.g., GPUs connected with NVLink, and machines connected with networks). This strategy partitions the deep model's operations into different workers (or stages) and has peer-to-peer communications between them. The batch data is partitioned into small micro-batches and injected into the pipeline, and the schedule plan for the pipeline parallelism plays an important role in maximizing throughput and memory savings. PipeDream\cite{pipedream} and DAPPLE\cite{dapple} adopt 1F1B schedule plan, which alternately executes the forward and backward stages of different micro-batches to maximize device utilization while keeping peak memory consumption constant. Unlike pipeline parallelism, SPMD parallelism\cite{gshard}\cite{gspmd} partitions operations along one or more dimensions on different devices. Collective communications will be inserted to preserve mathematical equivalence. Modern distribution strategy often employ a combination of these parallelisms when scaling large models to thousands of devices.

When training large models on dedicated clusters, pipeline parallelism can perform ideally. However, this is not the case when using cloud platforms, where state-of-the-art distribution strategies with pipeline parallelism face one challenge: preempted network resources between stages can significantly reduce throughput due to the features of cloud-based networks. Their dedicated clusters are not conducive to full resource utilization, and it is difficult to isolate networks for specific training tasks on the software level. Furthermore, the input training is transmitted through network overlaps with the training process itself, further preempting communications introduced by pipeline parallelization. Thus, even though the data transfer between stages may be small, there is a high probability of heavy cross-stage communications in a preempted environment, common on cloud platforms.

We analyzed and found that the throughput can be improved in the preempted network environment by adjusting the schedule plan. Since both forward stage and backward stage accepts activations produced by their dependent stage as inputs, the start time of each stage may be affected by its inputs arrival time. The default 1F1B scheduler alternately schedules one forward stage and one backward stage to orchestrate execution. The bottleneck happens either the alternative forward or backward stalls due to their slow transmission of inputs, which introduces more bubble time. To reduce this, the 1F1B plan should be extended to accomodate to the unstable network. By scheduling another stage whose inputs are ready, rather than waiting for the stage whose inputs are stalled, the device efficiency can be improved by allowing for computation to overlap with data transfer.
We propose \OurSchedule schedule plan for this purpose, where k micro-batches of forward stages and the same of backward stages are taken as one schedule unit. Compared to 1F1B plan, it alternatively schedules $k$ forward stages and $k$ backward stages. The early backward principle is still followed between different schedule units, and $k=1$ is the special case of \OurSchedule. Despite the increased potential for overlapping communication and computation in a congested network as the value of $k$ increases, this comes at the expense of an increased memory burden. When increasing $k$ to micro-batch number, the schedule plan is equivalent to GPipe's, preventing the early release for memory on device. Therefore, considering the fixed  micro-batch size, selecting an appropriate value of $k$ enables each device to retain a sufficient amount of computation tasks to prevent blocking caused by communication, while  trade-offs between peak memory usage and computation performance, as 1F1B facilitates early backward scheduling. Additionally, model users always provides global batch size, and the divided micro-batch size can be variant. Consequently, the optimal $k$ value should be searched for each micro-batch size. However, using a smaller micro-batch size may lead to a lower computation efficiency, which can decrease the end-to-end performance, despite the potential benefit of overlapping communication and computation from a larger $k$. 
Furthermore, the superior performance may not be sustained when the network traffic fluctuates. The schedule plan must be altered regularly in response to fluctuations in network traffic.

We present \OurApproach, an auto-adaptive scheduler designed with the following guiding principles:

$\triangleright$ \OurApproach works on the task graph that consists of stage computation instances. In our system, each stage computation is partitioned from HLO module\cite{jax}, a compiler intermediate representation (IR) used by
XLA\cite{xla}. Each stage computation is compiled and used to create execution instances known as task nodes, which \OurApproach schedules for execution on cluster.

$\triangleright$ \OurApproach adopts \OurSchedule scheduling, taking $k$ micro-batches of forward stages or backward stages as one schedule unit, with cross stage communications triggered immediately after each stage computation delivers its outputs. This design provides more opportunities for overlapping the computation and communications, while also reducing peak memory usage by releasing memory footprint immediately when no longer referenced.

$\triangleright$ \OurApproach regularly and automatically tunes $k$, recognizing that a larger k requires smaller micro-batch size, which might cause device underutilization. Taking into account the computation efficiency and frequent network preemption changes within a cloud platform, it is not practical to keep a fixed k from the beginning to the end. Thus, \OurApproach makes multiple versions of scheduling plans for different $k$ values and automatically tunes online to select the best version.

\OurApproach is implemented based on \PreviousWork\cite{zhang2022accelerating}, which is an automatic parallelization system. Our experiment demonstrates that our approach has the capability to adaptive to the preempted network environment, yielding stable performance. Compared with 1F1B schedule plan, \OurApproach speed training up from 4\% to 30\% in GPT and Unet.

\section{Backgroud and motivation}
\subsection{Parallelizations on Large Model}

Modern deep learning frameworks treats the process of deep learning computation as dataflow graph\cite{tensorflow, jax,pytorch}, with model developers building models through the use of fine-grained operators that are represented as nodes and connected with multi-dimensional tensors as edges. To facilitate optimization, some frameworks provide more customized or general intermediate representations (IRs), such as HLO in XLA\cite{xla}, JaxPR in \cite{jax} and MLIR\cite{lattner2020mlir}, which enable general and systematic distributed optimization in the compilation phase\cite{dapple,gshard,gspmd}. After compilation, the execution engine executes the plan on cluster.

When a large model is unable to complete training on a single device due to memory constraints or performance requirements, distribution strategies are employed to parallelize the model. Data parallelism\cite{sergeev2018horovod} is the easiest pattern to implement on most deep learning frameworks which involves splitting the training data across multiple devices and performing gradient reduction prior to parameters updates. Operator parallelism\cite{gspmd,gshard,lu2017flexflow} generalizes from data parallelism, allowing for splitting on any operator and requiring resharding communications to maintain correct computation. Both data parallelism and operator parallelism can be categorized as intra-op parallelism\cite{zheng2022alpa}. Pipeline parallelism\cite{gpipe,dapple,pipedream} is an inter-op parallelism that groups operations referred to as stages from the model graph and places them on different devices. All of these parallelization techniques are employed to maximize hardware resource utilization.

\subsection{Pipeline Parallelism}
Pipeline parallelism exhibits excellent performance on heterogeneous clusters connected by low-bandwidth networks across machines. In practice, one should find the cut-point, aiming to minimize the cross-stage communications while maintaining balanced stage computations. In pipeline parallelism, throughput performance can be measured by the number of bubbles produced. For synchronous training, injecting more micro-batches is a common method to reduce bubbles, although they cannot be eliminated completely. Theoretically, the throughput will eventually converge to the upper bound\cite{dapple}. Memory constraints are also an important factor, limiting the number of micro-batches that can be pushed in the warm-up phase. With the help of memory optimization techniques such as gradient checkpointing\cite{feng2021optimal, chen2016training}, more micro-batches can be injected into the pipeline, further reducing bubbles.

\subsection{Early Backward Scheduling}
Scheduling is essential in pipeline parallelism and must be carefully designed. Previous research has shown that scheduling affects both throughput and peak memory usage. GPipe injects micro-batches sequentially, causing an excessive memory issue on the first-stage device, as the activations produced by the forward stage of each micro-batch must be preserved until its backward phase is completed. Recent studies have improved the scheduling to form a 1F1B order in both synchronous\cite{dapple} and asynchronous training\cite{pipedream}, guided by the principle of early backward. This aims to release the forward activations on each stage as early as possible, allowing for more micro-batches or larger micro-batch sizes.  It has been demonstrated that, in a stable exclusive network environment, 1F1B ensures that the number of bubbles remains unchanged while keeping the peak memory usage constant in the pipeline's stable phase. Nevertheless, the number of bubbles drastically increases when the network is unstable.

\subsection{HLO and Task Graph}
To provide support for different frontend platforms and backend hardware, AI compilers (such as XLA\cite{xla} and TVM\cite{chen2018tvm}) typically introduce platform- and hardware-agnostic IRs for tensor programs. Specifically, the XLA compiler translates user-defined models into HLO IR and performs optimization passes. In our previous work\cite{zhang2023auto}, we developed an automatic parallelization pass for exploring distribution strategies, including pipeline parallelism. It analyses the computation graph formatted in HLO IR by minimizing cross-stage communication cost and decomposing the HLO graph into stage computations. We treat each HLO stage computation as a program and create its corresponding running instance called a task node. Task nodes fed by different micro-batches, but belonging to the same stage, share the same HLO stage computation. We create gradient accumulation task nodes for stitching different micro-batches. All task nodes are connected according to the data dependencies. The task graph is  automatically constructed, which is the description of the distribution strategy over the original deep model. To accommodate device assignment, Send/Recv pairs must be inserted. To achieve this, a special kind of task node is built and automatically inserted into the task graph, representing send and receive for peer-to-peer communication. In our work, the scheduling plan is created from the task graph. Since the task graph's granularity is relatively coarse, it significantly reduces the time spent generating the scheduling plan.

\subsection{Opportunities}
To accelerate synchronous pipeline parallel training, existing proposals focus on improving performance in exclusive network which are not feasible on cloud platforms\cite{rasley2020deepspeed,dapple,shoeybi2019megatron} due to the fragmentation of resources on public cloud services. Setting aside a network-isolated dedicated cluster is costly and not suitable for industrial use.  As a result, distributed GPU tasks are likely to be assigned to machines with preempted network environments, including data transfer or other non-GPU distributed tasks. This can lead to a decrease in performance, as the network resources are occupied by other tasks and delay cross stage communications.

The existing 1F1B scheduling behaves outstanding on dedicated cluster, but its performance is severely poor in a preempted network environment. It introduces more bubbles since the network resources are occupied by other tasks, resulting in the cross stage communications becoming non-negligible. Fig.~\ref{fig:kfkb}(a) demonstrates that the start time for each stage computation may be delayed due to the non-negligible cross stage data transfer, consequently leading to an increase in bubble time in the pipeline's stable phase. These bubbles, caused by scheduling the unready stage which is waiting for its inputs from the network, cannot be eliminated even with more micro-batches being injected. Previous literature has suggested that data transfer can be treated as an independent stage\cite{dapple} and that scheduling and injected micro-batch numbers can be re-planned accordingly. However, this approach is not practical, as the network resources between stages can vary greatly, making it difficult to hide communications via 1F1B scheduling, resulting in more bubbles.

Existing approach fails to consider the dynamic nature of network resource usage will change in real time. Consequently, fixed schedule planning is unable to adjust to the changing environment. For instance, if the network resources between two stages are periodically occupied by other tasks, the fixed schedule plan may not suffice, even if we take the data transfer stage into account as non-negligible and independent stage. Moreover, the variations in network resource usage between different stages make it difficult to plan for the future.

If we revert back to the GPipe case for 1F1B scheduling, we speculate and observe that the forward and backward stages both overlap with communication more efficiently in a preempted network. The injection of forward and backward stages sequentially provides more opportunities for overlap. For example, for forward stages, the micro-batch $n-1$ in stage $s$ can overlap with the delivery of outputs of micro-batch $n$ in stage $s+1$ when $n  \textgreater 1$. thus, reducing the bubble time compared to 1F1B scheduling. Nevertheless, GPipe's scheduling can lead to a greater peak memory usage.

There is a trade-off between pipeline efficiency and peak memory usage in preempted network environment when adjusting schedule plan to \OurSchedule style. Consequently, we advocate more general and adaptive tuning of schedule plans, striking a suitable balance between computation efficiency and memory usage. 

\subsection{Challenges}
Exploring the opportunities comes with challenges.

\noindent\textbf{Time to evaluate search candidates.} For \OurSchedule scheduling, $k$ forward or backward stages serve as basic schedule units. The group member count $k$ must be decided, yet it can also be affected by micro-batch size $b$ given the limited device memory. The pipeline length should be estimated based on stage execution time and network profiling results, as there are many feasible combinations, yet the evaluation time spent for the pipeline length in an ever-changing network environment is unacceptable due to the need for profiling between each stage. Obviously, pruning the candidates set is necessary to reduce the evaluation time. To address this problem, an efficient algorithm is designed to effectively prune the candidate set on the Pareto optimal frontier.

\noindent\textbf{Difficulty in accurate execution time prediction of pipeline length.} To estimate the pipeline length, we should profile both stage execution time and cross-stage communication time. While stage execution time can be profiled accurately, cross-stage communication time is more difficult due to the varying network occupancy between host machines and the time-sensitive nature of the network occupancy. and the lack of proportionality between stable network environments and data size. Furthermore, even if the network is stable, the cross-stage communication time will not be proportional to the data size. To address these challenges, we propose an adaptive practical profiling strategy to accommodate the ever-changing network environment.

\section{Overview}

\begin{figure}[t]
    \centering
    \includegraphics[width=0.51\textwidth]{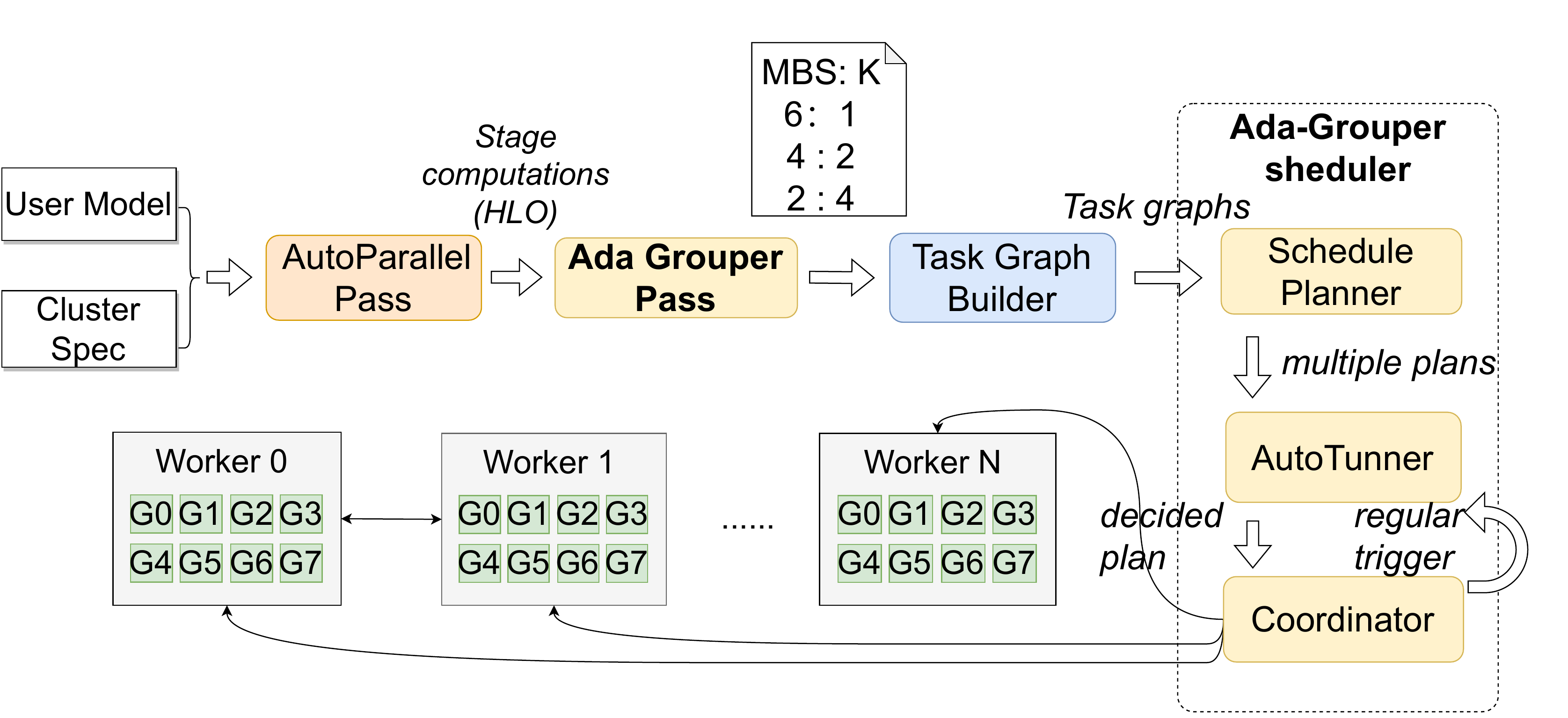}
    \caption{\OurApproach workflow\label{fig:overview}}
\end{figure}

\OurApproach is a standalone component developed based on our previous work \PreviousWork\cite{zhang2023auto}. As illustrated in Fig.~\ref{fig:overview}, the end-to-end high-level workflow of \OurApproach which features an \emph{\OurApproach pass} and an \emph{\OurApproach scheduler}. The \PreviousWork's automatic parallelization pass takes a user model and cluster specification as inputs to explore the SPMD and pipeline parallelization space and generate a set of HLO stage computations, which serves as the precursor to our work. With \OurApproach, $k$ forward or backward stages are combined into a single group and scheduled as an indivisible unit. To determine the optimal value for group member count $k$ given the various micro-batch sizes that can be accommodated under the memory limit, an enumeration of all possible solutions is required. As such, the \OurApproach pass searches multiple versions of micro-batch size ($b$) and group member count ($k$) pairs. It also provides an efficient pruning algorithm to enumerate all potential optimal candidates.

Next, the task graph builder from \PreviousWork compiles all stage computations in the given set of candidates to generate respective execution binaries, subsequently constructing multiple versions of task graphs according to the its specified group member count and micro-batch size pair. The \emph{Ada-Grouper scheduler} takes these task graphs as inputs and schedules adaptively. Specifically, the \emph{schedule planner} make \OurSchedule scheduling plans according to the group member count ($k$) and micro-batch size ($b$) values for each task graph as candidate plans. The \emph{auto tunner} evaluates all candidate plans and selected the optimal one by using a cost model that estimates the pipeline length through profiling the network and computing the execution time of each stage. Other candidate plans are preserved for later selection. Then, the coordinator orchestrates different workers according to the decided plan. To adapt to the ever-changing network environment, \OurApproach regularly triggers \emph{auto tunner} to re-evaluate each candidate, selecting a new plan which is optimal in the future but with a potentially different $k$ value. When the plan is decided, it takes effect immediately upon notification of the policy change.

\subsection{\OurApproach Pass}
\OurApproach pass enables the selection of multiple candidates at different $k$ values, as each candidate has the potential to be replaced in a dynamic network environment. It is noted that 1F1B is the most memory-efficient, as the early backward process promotes the early release of corresponding forward activations\cite{dapple}, and larger $k$ value consumes more memory. To ensure the generation of candidates that are within the memory limit, the memory usage of each plan generated from group count and micro-batch size pair is estimated while enumerating $k$ in the current phase. Obviously, a larger $k$ value is always paired with a smaller $b$. The smaller micro-batch size does not have better computation efficiency, although it may bring better overlapping with communication to with the combination of larger group count. This trade-off between computation efficiency and pipeline throughput in runtime is taken into account through the effective search algorithm designed for this pass, allowing for the generation of all necessary potential candidates.

\subsection{\OurApproach Scheduler}
\subsubsection{Schedule Planner}
The schedule planner takes over all task graphs constructed by task graph builder and makes \OurSchedule plan. Utilizing the principle of early backward, the planner schedules one backward group as soon as possible to minimize peak Memory usage. The \OurSchedule algorithm works on each version of the task graph, preserving all task graphs and their corresponding schedule plan permanently, without deciding which Plan is to be executed.

\subsubsection{Auto Tunner and Coordinator}
\emph{Auto tunner} regularly decided one group schedule plan to adapt to the ever-changing preempted network. The trade-off between computation efficiency and pipeline throughput implies that each solution has the potential to be the optimal. To achieve this, a simple cost model is designed to evaluate pipeline length under current network at regular intervals. This cost model requires profiling data of stage computation time and cross stage communication time, which need to be executed multiple times in order to produce more reliable results. It should be noted that the computation time cost is stable during training since there is no preemption to computation device, thus its profiling data can be reused. However, network preemption makes re-evaluation necessary. The \emph{coordinator} is setup for parallel execution in runtime, scheduling each task node on its device via multi-threading and dispatching the newly decided schedule plan to all participant workers. After the auto tunner chooses the suitable plan, the coordinator will modify the schedule plan and take it into effect immediately.

\section{Group Schedule Plan}

\begin{figure*}[ht]
    \vskip -1em
    \centering
    \includegraphics[width=0.99\textwidth]{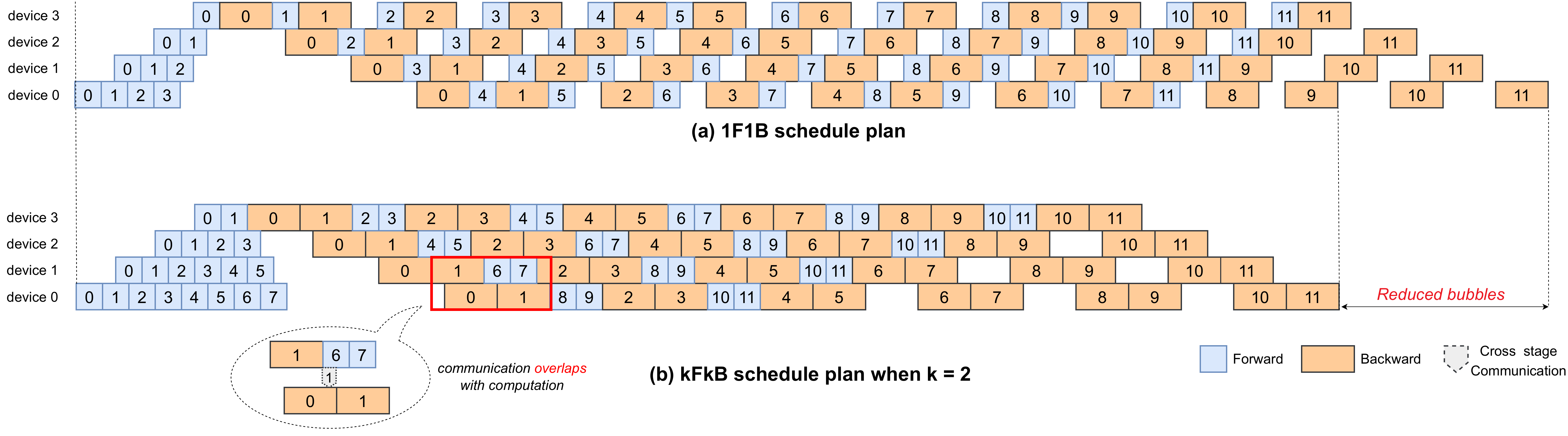}
    \caption{Pipeline length analysis of 1F1B and 2F2B schedule plan in preempted network environment.\label{fig:kfkb}}
    \vspace{-1em}
\end{figure*}

\subsection{kFkB Pipeline Length}
\label{sec:kfkb_pipeline}
We reveal the pipeline length comparison between \OurSchedule and 1F1B schedule plan in preempted network environment. We make two reasonable assumptions: (1) the time cost of backward stage is almost double of that of the forward stage; (2) the time required for cross-stage transfer time is half of the forward stage computation time.

Fig.~\ref{fig:kfkb}(a) demonstrates that the 1F1B schedule plan significantly worsens the performance in this situation. The time taken for data transfer between stages is non-negligible. Theoretically, the proportion of bubbles introduced by 1F1B plan is higher than that in an exclusive network environment. In comparison, Fig.~\ref{fig:kfkb} reveals that the \OurSchedule plan yields shorter pipeline length when $k \textgreater 1$.

Analytically, in the 1F1B scenario, more non-negligible bubbles are generated when the network is occupied. The cross-stage communication in such a network environment impedes task scheduling. This deteriorating phenomenon can be alleviated as long as the network preemption is not serious.
However, on cloud platforms, achieving an ideal environment is a rarity. When increasing $k$ from 1 to 2, the overall throughput is significantly improved. Zooming in on the \OurSchedule pipeline, we observe that after combining two forward stages or two backward stages into one scheduled unit, the second computation overlaps with the cross stage communication followed by the first computation, thus preventing each device from being blocked due to cross-stage communication by introducing another ready task. The more opportunities for computation and communication overlaps, the more bubbles are eliminated. We can speculate that the proportion of bubbles can be further decreased when pushing $k$ from 2 to larger.

Another point to be aware of is that increasing the $k$ value will bring side effect on memory usage. If  $k$ is set to $M$(the number of micro-batches), the schedule plan reverts to that of GPipe\cite{gpipe} case.  This presents an overwhelming amount of memory pressure on the device where the first stage is located, since its activations must be preserved in memory almost throughout the entire training iteration. A potential solution to alleviate this issue is to further divide the micro-batch along batch dimensions. However, this may reduce computational efficiency since the micro-batch size would be smaller.

\subsection{Pruning for Candidate Set}

In order to generate \OurSchedule schedule plan, we should generate all available solution candidates. The \OurSchedule schedule plan is related to two variables: the basic schedule unit measured by group member count $k$, which provides the opportunity for overlapping computations and cross-stage communications, and the micro-batch size $b$, which is key to the computation efficiency. In principle, the solution is available and will be placed in candidates set if it satisfies memory constraint. Then, \OurApproach evaluates pipeline length for each schedule plan and selects the best one. However, the evaluation is time-consuming due to need for network profiling. For example, two schedule plans with different micro-batch sizes may have completely different communication performances. Additionally, communication performance should be profiled in the ever-changing network environment. If the evaluation time is too long, there is a high probability that the evaluation will be invalid as the network environment has already changed. To address this issue, we designed a pruning algorithm that significantly reduces the candidate set, which helps to alleviate the evaluation pressure in runtime.

\begin{figure}[ht]
    \centering
    \includegraphics[width=0.45\textwidth]{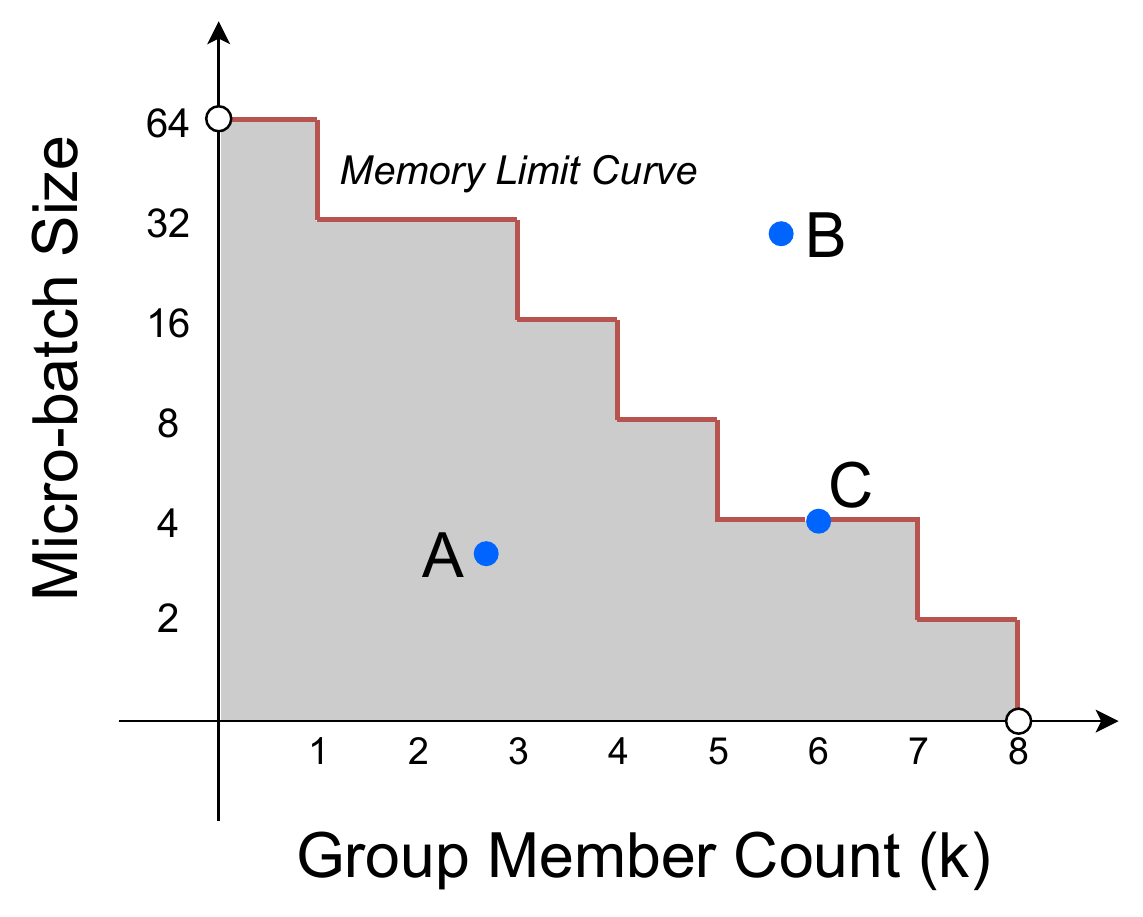}
    \caption{Generate candidates on memory limit curve. The shadow area where solution $A$ located is available but has the lower memory utilization. The area where solution $B$ located causes out of memory. All combinations of integers on the memory limit curve are our candidates. \label{fig:enumeration_curve}}
\end{figure}

Fig~.\ref{fig:enumeration_curve} shows our candidate sets and essentially it is the Pareto optimal frontier. Increasing the group member count ($k$) or micro-batch size ($b$) consumes more memory until the device memory bound is met. By fixing $k$, we can easily obtain the maximum micro-batch size and these combinations form the memory limit curve. All the combinations on or under the memory limit curve are available solutions, but the under-utilized memory solutions should be pruned. The point $B$ over the memory limit curve cannot be chosen for memory constraint. However, if we choose a combination like point $A$ in shadow area, the device memory is under-utilized. Therefore, we need to pay attention only to the points on memory limit curve like $C$, which helps to drastically prune the candidate set. Specifically, we can start by gradually increasing the group member count $k$ from 1 and then greedily search for the maximum micro-batch size that can be accepted.

\subsection{Cost Model}
\OurApproach utilizes a cost model which requires stage computation and cross-stage communication profiling data to estimate pipeline length for all available schedule plans.The stage computation time cost is stable due to its exclusive assignment to the designated device and no need to be re-evaluated during the training process. Notwithstanding, the network performance should be monitored periodically since it tends to fluctuate. Thus, multiple profiling actions should be conducted during a period of time and the moving average of these results should be taken as the final outcome.

Instead of estimating the time cost of data transfer by measuring the bandwidth between stages, we measure the cross-stage communication time directly. This approach is favored for two reasons: first, although the same amount of data is transferred between stages, the performance may vary significantly depending on the severity of preemption; second, the bandwidth utilization of tensors with different shapes and sizes is not the same. Thus, profiling the end-to-end cross-stage communication provides more reliable data and is more efficient. Furthermore, the profile data for each scheduling plan should be averaged in a window at a certain interval.

\subsection{Analysis in Unstable Network}
\begin{figure}[ht]
    \centering
    \subfloat[Part of pipeline parallelism in unstable network] {
    \label{fig:unstable_bubble}
    \includegraphics[width=0.45\textwidth]{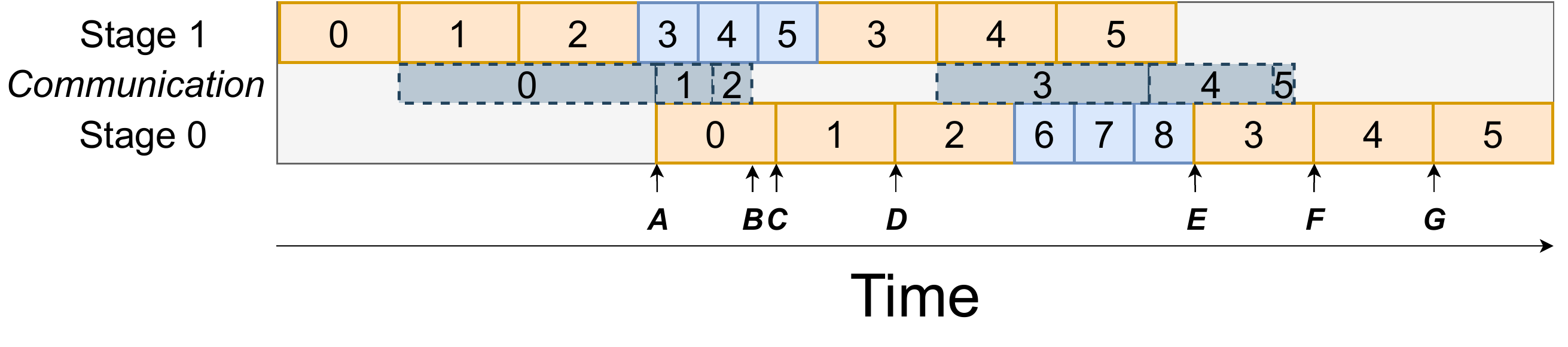}}
    \newline
    \subfloat[Cross stage bandwidth for each micro-batch in \ref{fig:unstable_bubble}] {
    \label{fig:bandwidth}    
    \includegraphics[width=0.20\textwidth]{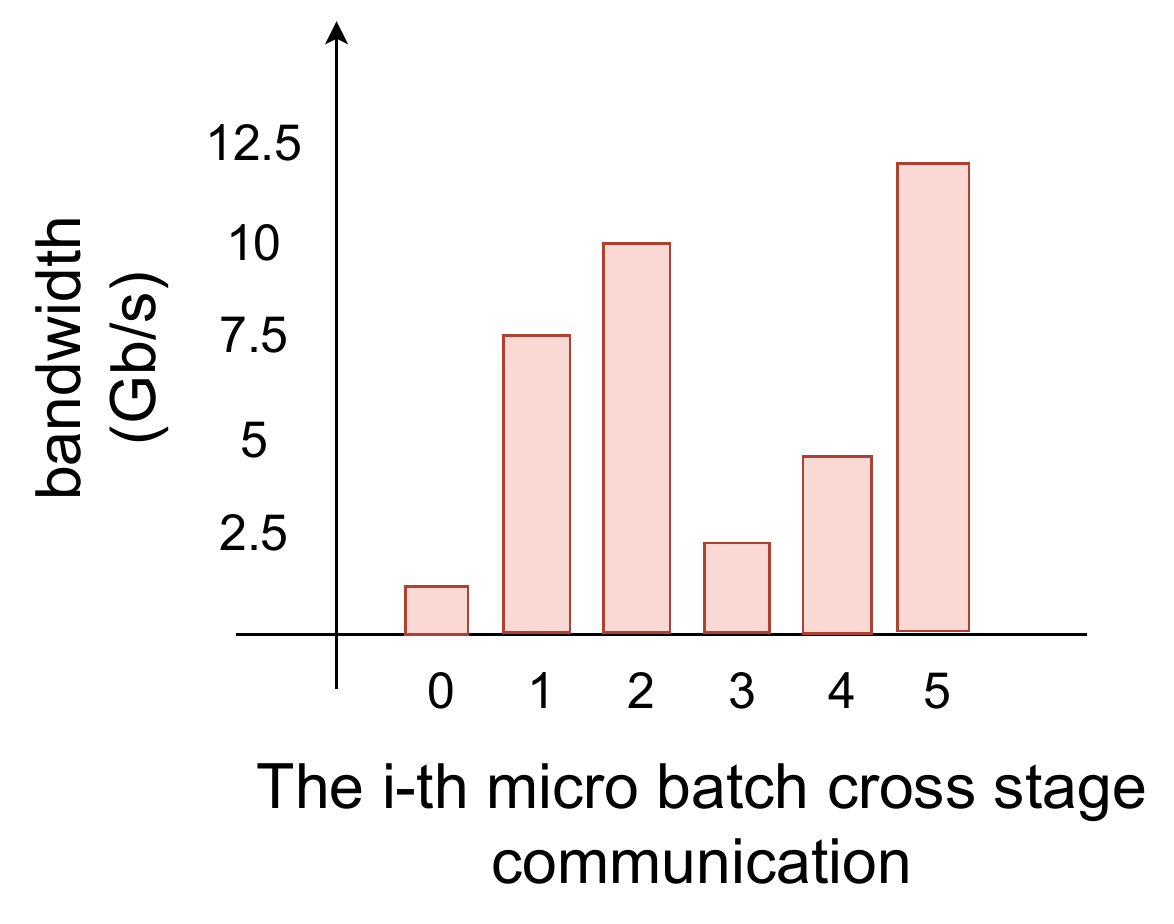}}
    \subfloat[The arrived inputs in buffer queue] {
    \label{fig:buffer_queue}    
    \includegraphics[width=0.25\textwidth]{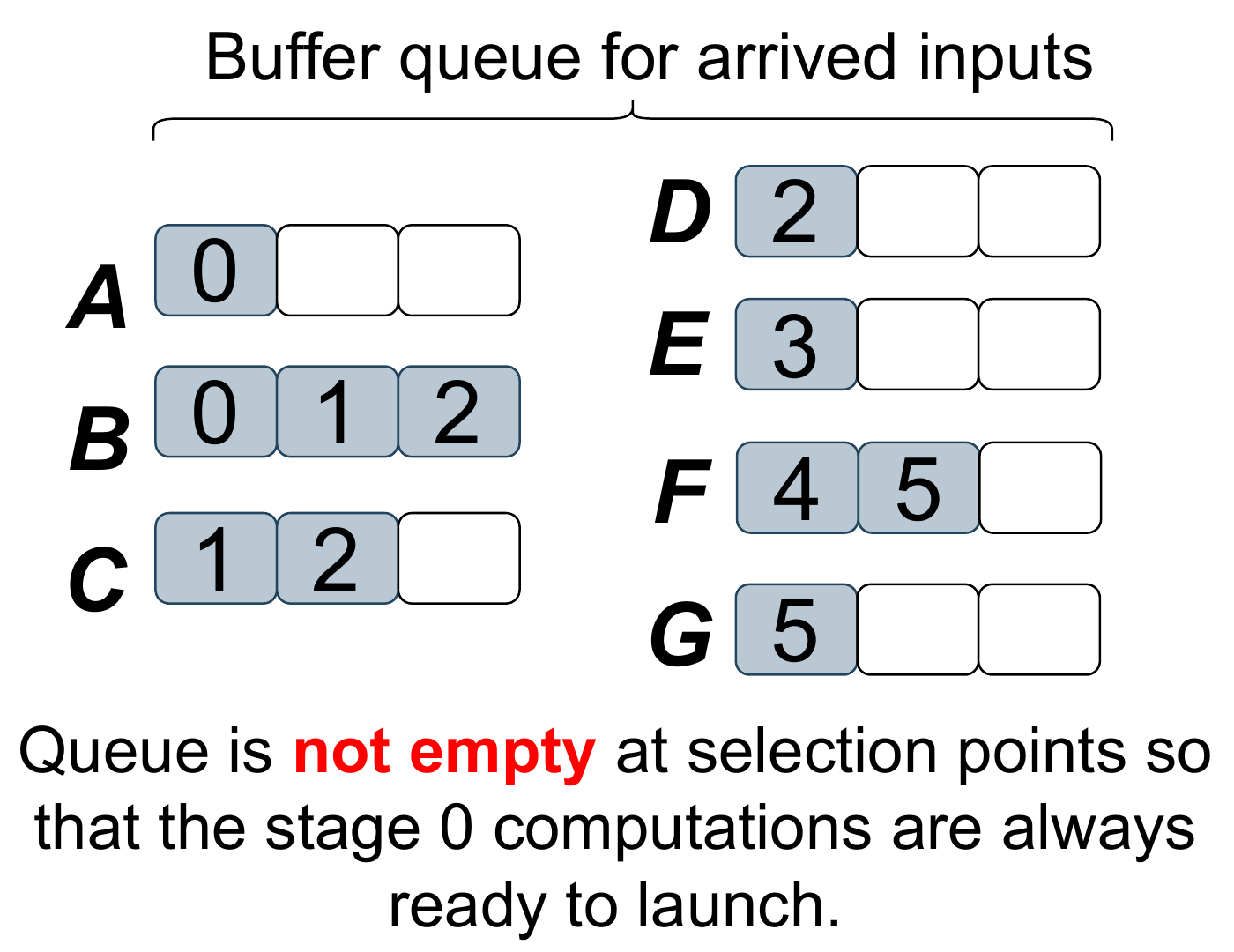}}
    \caption{The 3F3B performance analysis in unstable network.}
\end{figure}

We explain the performance stability in unstable network by using 3F3B schedule plan. As shown in Fig.~\ref{fig:unstable_bubble}, the 3F3B pipeline is affected by the instability of the network. The micro-batch 0, 1 and 2 experience backward, leading to the communication from stage 1 towards stage 0. The sudden fluctuations in bandwidth between stage 1 and stage 0 lead to poor and unstable communication for each micro-batch in backward computation flow. Fig.~\ref{fig:bandwidth} illustrates the cross-stage communication bandwidth between the two stages for each micro-batch. The low bandwidth for micro-batch 0 delays the start time of corresponding stage computation in stage 0. However, the network state for micro-batch 3 is also poor but has no negative impact on its following stage computation. 

To explain this phenomenon, it is assumed that there is a buffer queue to store the cross-stage communications from stage 1 to stage 0. When a computation in stage 0 is completed, the liveness of its inputs is terminated, enabling it to be popped from the queue. Consequently, the buffer queue stores the ready inputs for upcomming stage computation in the present moment. For the computation to proceed without being postponed when preparing for launch, the queue must not be empty. The inputs are sent beforehand regardless of the long transmission time. Seven points ($A$ to $G$) in Fig.~\ref{fig:unstable_bubble} are selected to check the queue state. Except for point $B$, all the selection points are the time for the launch of computations and there queue states\ref{fig:buffer_queue} are not empty, which indicates that their required inputs have arrived ahead of time. As for point $B$, the queue is full due to the fact that computation for micro-batch 0 has not been completed, thus it is unable to pop its inputs. The most concerning point is $E$. Even though its input transmission was affected by network preemption, resulting in a longer transmission time, similar to the inputs of point $A$, its computation is not postponed. In 3F3B scheduling, point $E$ is the earliest opportunity for launching backward computation of micro-batch 3 in stage 0, depending on whether its inputs has arrived in advance. We found that the queue stage of $E$ is not empty and its inputs already presents in the queue. This was the fundamental reason why the pipeline did not generate additional bubbles when the network was preempted yet again. Overall, when the network is in a high-speed state, an increased number of arrival inputs will be pushed in the queue, providing sufficient resources to be consumed when the future network is preempted. This advantage enables the pipeline system to maintain a good throughput performance, even in an extremely unsteady network state.

\section{Implementation}

We design and implement \OurApproach based on our previous work \PreviousWork, which parallelizes user models on HLO modules to construct a task graph and schedule it on distributed devices. The core of the \OurApproach pass, which is placed after AutoParallel pass(as shown in Fig~.\ref{fig:overview}) is an effective enumeration algorithm that generate sufficient candidates. The scheduler planner in the \OurApproach scheduler is modified from 1F1B scheduler. We generate $k$ copies of 1F1B scheduling sequences and interleave them to build a \OurSchedule plan, saving all available plans for future selection. The cost model plays the important role in the auto-tunner and coordinator, as it accurately evaluates the pipeline length via the available profiler network periodically. Finally, the cross-stage peer-to-peer communications should be launched asynchronously launched and should not block device or host. We have implemented dedicated task nodes(Send/Recv pairs) for NCCL primitives\cite{nccl} to handle peer-to-peer cross-stage communication. 

Section \ref{sec:micro-batch-grouper-pass} presents how to enumerate candidates in \OurApproach pass. Section \ref{sec:profiling} discusses the cost model details and profiling strategy. Section \ref{sec:asynchronous-p2p} shows our asynchronous design for peer-to-peer communication. Section \ref{sec:online_tuning} simply introduces the \OurSchedule scheduling and online tuning implementation.

\subsection{\OurApproach Pass}
\label{sec:micro-batch-grouper-pass}
We have developed the \OurApproach pass, which is placed after our AutoParallel pass, to enumerate and prune the set of candidates. As inputs, the pass takes the stage computations annotated with the batch size dimension and decomposed into stages automatically. During the enumeration process, it splits the stage computation across the batch size dimension via operator splitting inference, while the Buffer Assignment module in XLA is used to accurately estimate the memory usage of the slimming HLO computation. After k and b are decided, the corresponding scheduling plan is also determined and evaluated using a simple memory cost model to ensure it satisfies the memory limit constraints, considering the liveness of the outputs of each stage computation.

\subsection{Profiling}
\label{sec:profiling}
\noindent\textbf{Stage execution time.} We profile stage execution times for each schedule plan individually. To ensure more reliable profile data, each stage computation needs to be measured multiple times and calculate the average. As all the assigned devices are guaranteed exclusive, there is no need to re-profile all stage execution times during the online tuning phase.

\noindent\textbf{Cross stage communication time.} We profile the cross-stage communication time between stages for all schedule plans under current network environment. Similar to the stage execution time, ach cross stage communication time should also be profiled multiple times and takes its average However, it requires to be re-profiled during online tuning phase for the network environment may have changed. To ensure the profiling data not to be disturbed by our own task, we suspend the current schedule task and collect all the performance data in each schedule plan.

\subsection{Asynchronous P2P Communication}
\label{sec:asynchronous-p2p}

\begin{figure}[ht]
    \centering
    \includegraphics[width=0.47\textwidth]{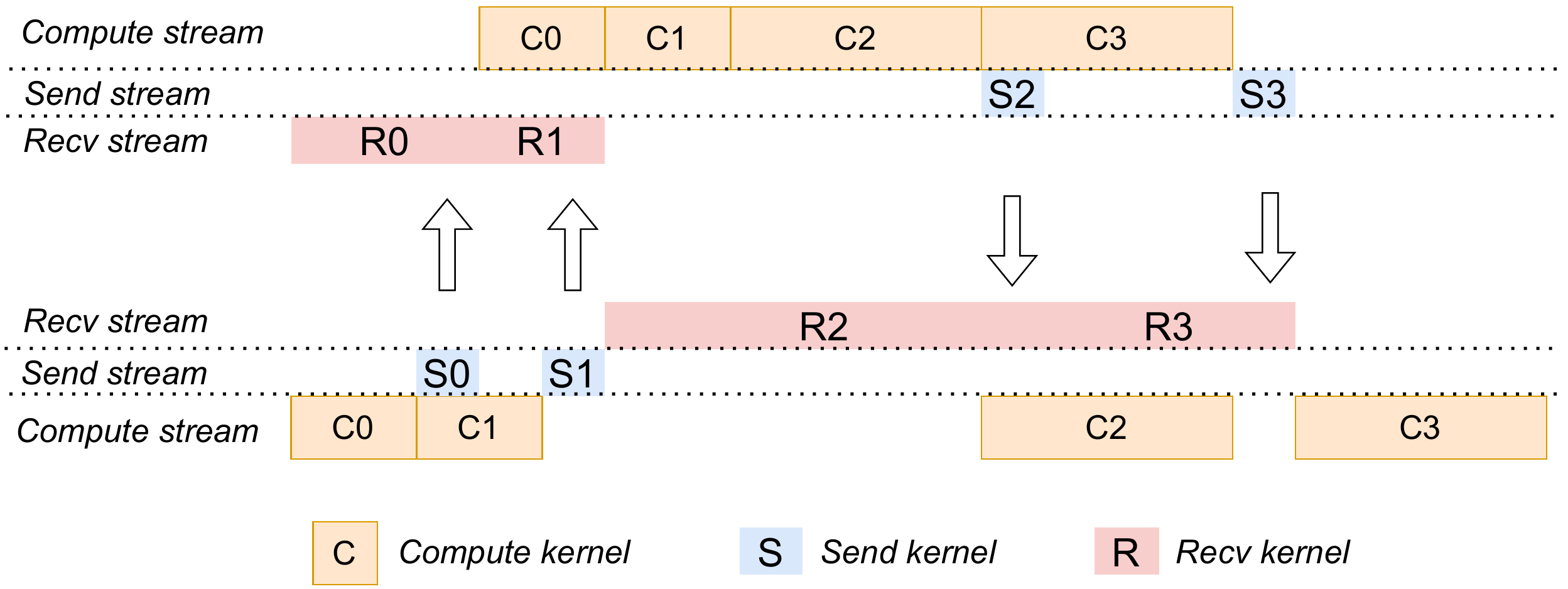}
    \caption{\OurApproach asynchronous p2p communication\label{fig:async_p2p}}
\end{figure}

The \OurSchedule schedule plan benefits from the overlapping between cross stage communication and stage computation, which necessitates an asynchronous form of peer-to-peer(p2p) communication. To this end, we design and implement the p2p communications using NCCL primitives guiding by three principles. First and foremost, the send kernel and receive kernel should be launched on separate streams that are distinct from all other kernel streams used for computations. Secondly, the send and receive for both participants must be properly paired across devices without mismatch, otherwise it could result in deadlock or unpredictable behavior. Lastly, when different pairs of participants work on the same device and do p2p communication in the same direction, the created communicators should be reused. This strategy prevents the need to create duplicate communicators. Fig~.\ref{fig:async_p2p} presents our p2p communication design in \OurApproach, wherein sending, receiving and computation kernels run on different streams, respectively. The pairing between sending and receiving kernel is established through the scheduling sequence on each device. Additionally, $R1$ is launched after $R0$ is completed to make sure that the data transfer is not aligned, and $R3$ and $R4$ follow a similar approach.

\subsection{kFkB Schedule and Online Tuning}
\label{sec:online_tuning}
\noindent\textbf{kFkB schedule.}
The heuristic 1F1B scheduling algorithm is implemented, generating k copies of the 1F1B plan.  These k plans are then cross-merged to build the merged plan, with adjustments applied to ensure pairing of the Send/Recv task nodes. Then the \OurApproach pass produces a set of available \OurSchedule scheduling plans which are preserved in the Ada-Grouper scheduler and re-evaluated at regular intervals through online tuning.

\noindent\textbf{Online tuning.} The online tuning procedure, controlled by an environment variable and triggered at regular intervals, re-profiles cross-stage communication times and evaluates the pipeline length for each schedule plan to determine the optimal one. Switching between schedule plans does not require variable buffers to be dumped out and restored in the new plan due to the variance of micro-batch size or group member count having no effect on model parameters, thus resulting in a switch with minimal overhead. The coordinator dispatches the best schedule plan immediately after the decision of online tuning.

\begin{table}
	\centering
	\caption{GPT model test configurations. Different configurations are used for weak scaling tests on 1/2/4/8 workers. "GPT-Medium" is used for strong scaling and granularity tests. Global batch size is set to 64 for strong and weak scaling and 192 for granularity tests. All tests use float16 precision.\label{tab:gpt_conf}}
	\resizebox{\columnwidth}{!}{
		\begin{tabular}{lccccccc}
			\toprule
			Config Name & $N_{params}$ & $N_{layers}$ & $D_{hidden} $ & $D_{ffn}$ & $N_{heads}$ & $D_{head}$ \\
			\midrule
			GPT-Medium & 350M & 24 & 1024 & 4096 & 16 & 64 \\
			GPT-Large & 760M & 24 & 1536 & 6144 & 16 & 96 \\
			GPT-XL & 1.3B & 24 & 2048 & 8192 & 32 & 64 \\
			GPT-2.7B & 2.7B & 32 & 2560 & 10240 & 32 & 80 \\
			\bottomrule
	\end{tabular}}
\end{table}

\begin{table}
	\centering
	\caption{U-Net model test configurations. Both configurations are used for weak scaling (by samples) tests. Single float precision is used for all tests. The error bars show the performance varying range of different steps.\label{tab:unet_conf}}
	\resizebox{0.7\columnwidth}{!}{
		\begin{tabular}{lccc}
			\toprule
			Config Name & $N_{params}$ & $N_{dims}$ & $D_{image\_size} $ \\
			\midrule
			UNet-Base & 32M & 64 & 32 \\
			UNet-Medium & 768M & 320 & 32 \\
			\bottomrule
	\end{tabular}}
\end{table}

\begin{figure}[t]
	\centering
	\includegraphics[width=0.48\textwidth]{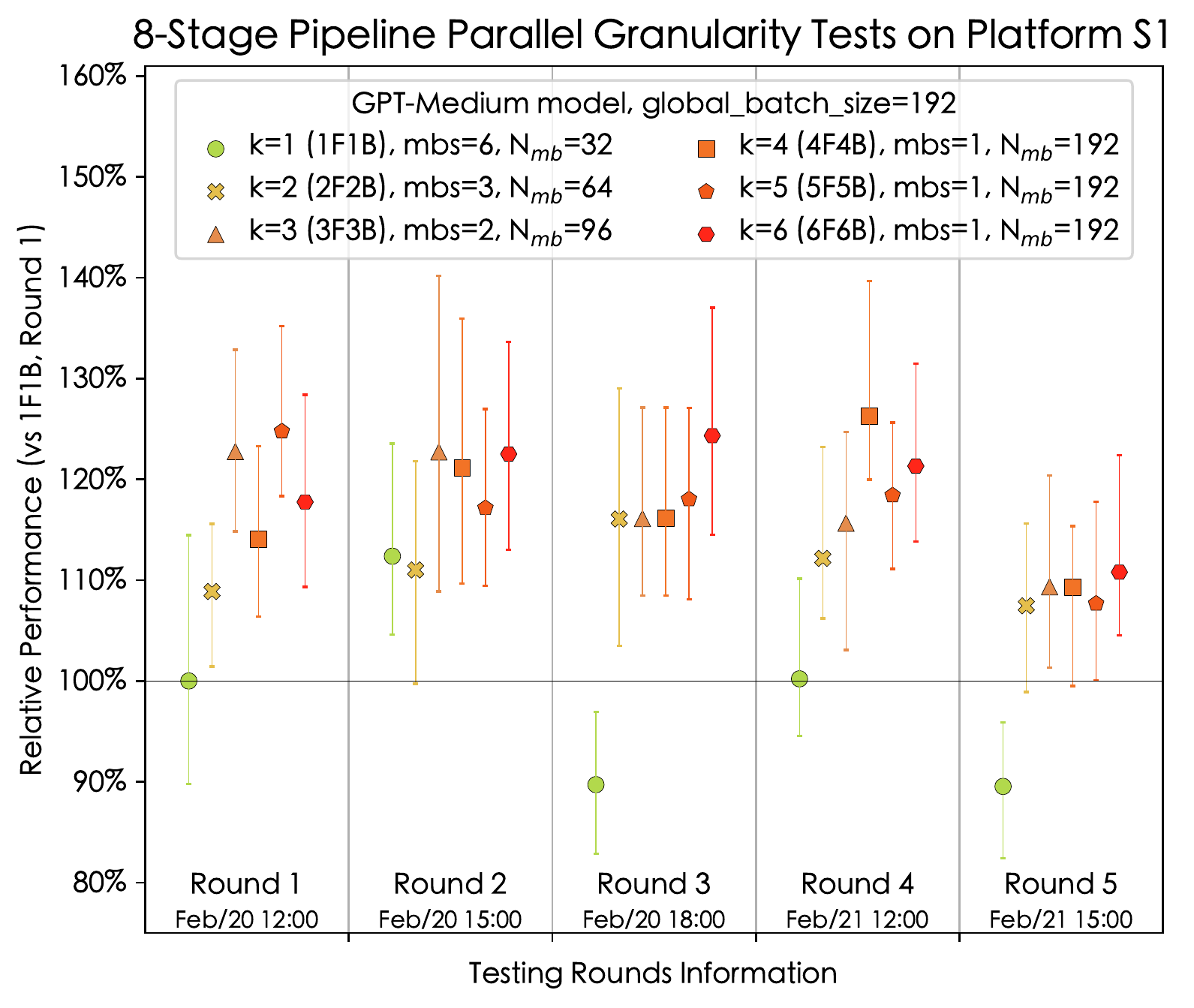}
	\caption{Pipeline granularity tests on 8 workers of Platform S1 with GPT-Medium configuration. All tests were done with the fixed global batch size 192, while micro batch size(mbs) and count(N$_{mb}$) were changed according various group scheduling count $k$ (from 1 to 6) of \OurSchedule. The error bars show the performance varying range of different steps.\label{fig:gpt_granularity}}
\end{figure}

\section{Evaluation}
\OurApproach is designed for adaptive micro batch scheduling by the \OurSchedule pipeline schema, which help making stage data transferring smoother when the network condition varies in the production environment. Thus we choose three platforms with different usage to experiment how \OurApproach could affect pipeline parallel model training workloads. In fact, it is not easy to precisely demonstrating the real time network condition in quantitative, thus the timings of multiple iteration steps would be expected varying in a range. The performance would also be affected while a test was done at the moment when the platform is relatively busy or free.

\subsection{Platforms and Models}
\textbf{Testbed}. The three platforms are all equipped with typical commodity hardware configuration. To be noted that 1) nodes in S1 and M8$_s$ are connected with the same network through which real production jobs are running, 2) workers on C$_{1x}$ and M8$_s$ might share the same physical machine resources with others workers, which belong to other jobs, or occasionally, the same job.  That means uncertain resource usage collision or network contention would sometimes occur, that should obviously influence to the tests performance. However, we'll consider that these situations make the test results more closer with the real world jobs.

\begin{itemize}
	\item \textbf{Platform C$_{1x}$:}: A cloud resource pool for internal tests. Configuration: (each instance) 1 * V100-SMX2-32GB GPU, 12 vCPU cores of Xeon (Skylake) 2.7GHz, 96GB DDR4-2666 vMem, 25Gb vEthernet.  
	\item \textbf{Platform S1:} An online exclusively developing platform. Configuration: (each machine) 1 * V100S-PCIE-32GB GPU, 2 * Xeon (Cascade Lake) 26C 2.5GHz, 512GB DDR4-2666, 1 * 100Gb RoCE interconnect across production environment switches.
	\item \textbf{Platform M8$_s$:} An online pre-production platform. Configuration: (each machine) 8 * V100-SMX2-32GB GPU w/NVLink2, 2 * Xeon (Skylake) CPU 24C 2.5GHz, 768GB DDR4-2666, 1 * 100Gb RoCE interconnect across production environment switches.
\end{itemize}

All machines are installed Linux OS with kernel version 4.19.91, NVidia driver 470.82 and NCCL library 2.8. The tests are run inside the developing docker equipped with CUDA 10.1 runtime and \PreviousWork parallel system\cite{zhang2023auto}.

\textbf{Benchmark models}. Two up-to-date high frequently used base models were chosen for \OurApproach evaluation experiments.  The first is GPT\cite{gpt3} model for NLP tasks, the other is U-Net\cite{unet} for text-to-image diffusion models. The
configurations used of the evaluation of the two models are listed in Table \ref{tab:gpt_conf} and Table \ref{tab:unet_conf}.

\textbf{Baseline}. Based on the same parallel training system \PreviousWork\cite{zhang2023auto}, the experiments compare \OurApproach with 1F1B pipeline scheduling results, on the platforms and models above. The proposed tests emphasize 1 GPU per worker cases, that would be much similar as cloud environment or multi-job online production system in which fragment resources are usually existing and allocated for jobs.

\begin{figure}[t]
	\centering
	\includegraphics[width=0.48\textwidth]{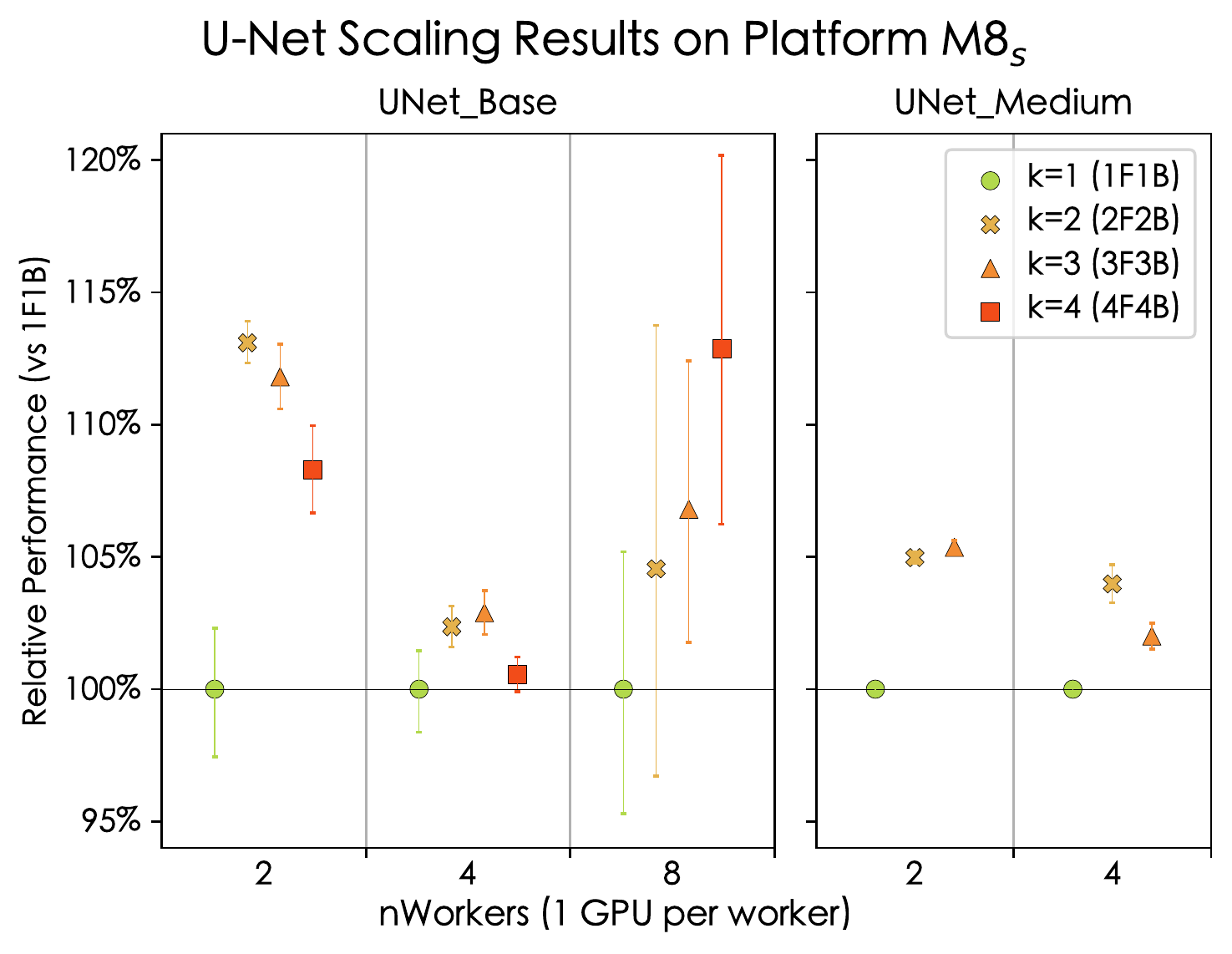}
	\caption{Relative performance comparison of different \OurSchedule on U-Net model, weak scaling by global batch size on Platform M8$_{s}$. Global batch size was setting as N$_{workers}$*128 for the tests. The error bars show the performance varying range of different steps.\label{fig:unet_weak}}
\end{figure}

\begin{figure*}[t]
	\centering
	\subfloat[] {
		\label{fig:gpt_weak_c1x}
		\includegraphics[width=0.33\textwidth]{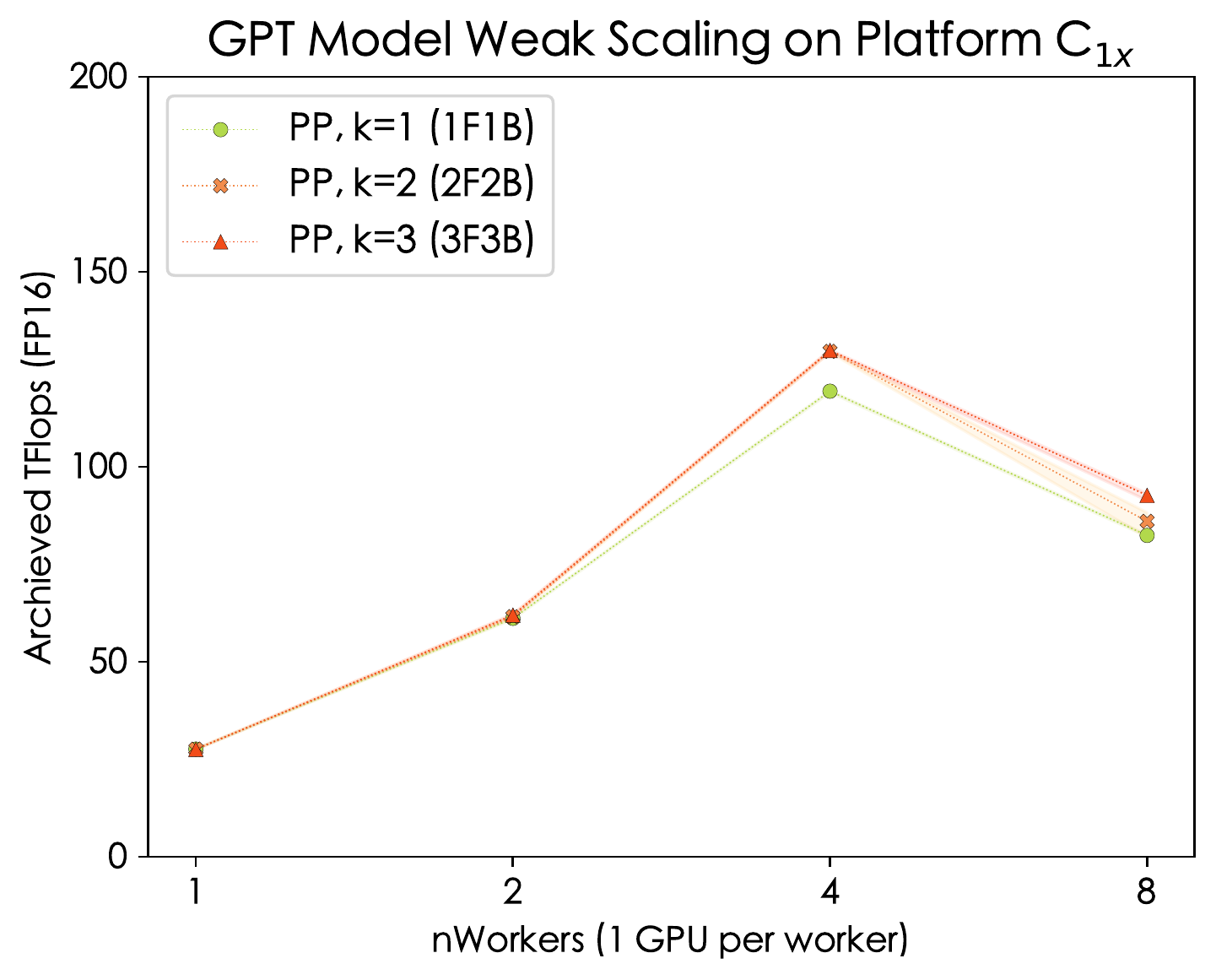}}
	\subfloat[] {
		\label{fig:gpt_weak_s1}
		\includegraphics[width=0.33\textwidth]{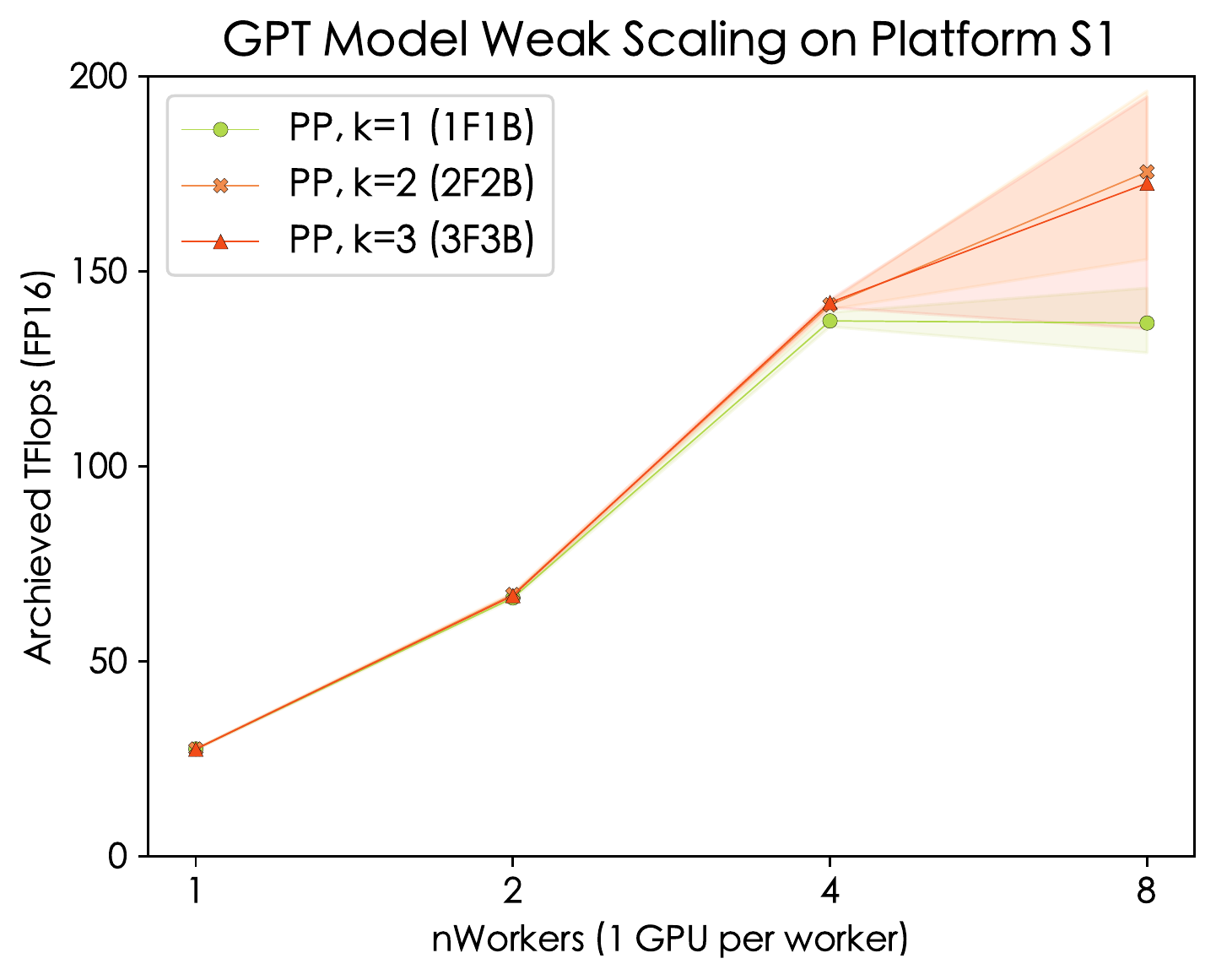}}
	\subfloat[] {
		\label{fig:gpt_weak_m8s}
		\includegraphics[width=0.33\textwidth]{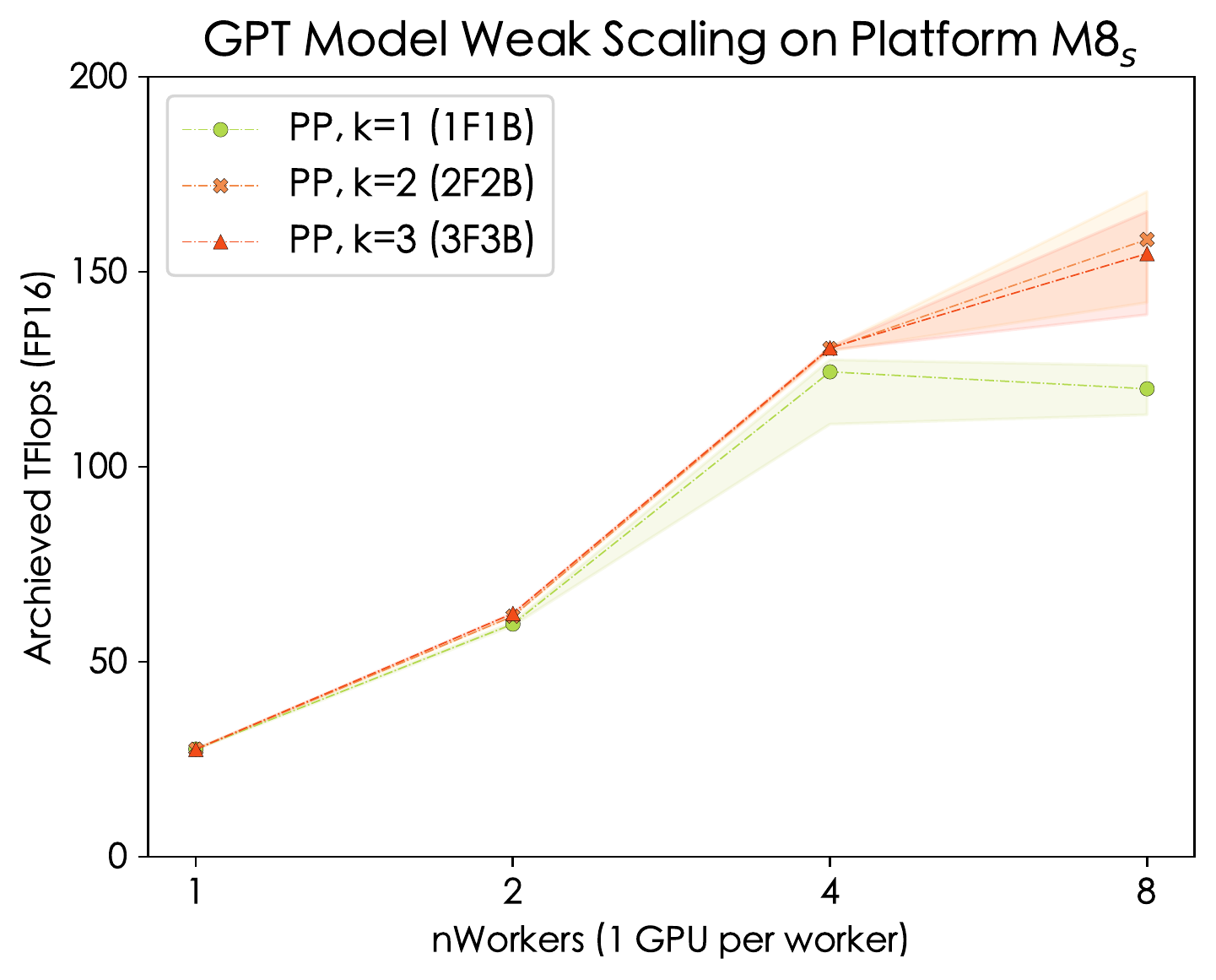}}
	\caption{GPT model pipeline parallel weak scaling (by arguments) performance. The shadow areas indicate the performance varying range of different steps. \label{fig:gpt_weak_tests}}
\end{figure*}

\begin{figure*}[t]
	\centering
	\includegraphics[width=0.85\textwidth]{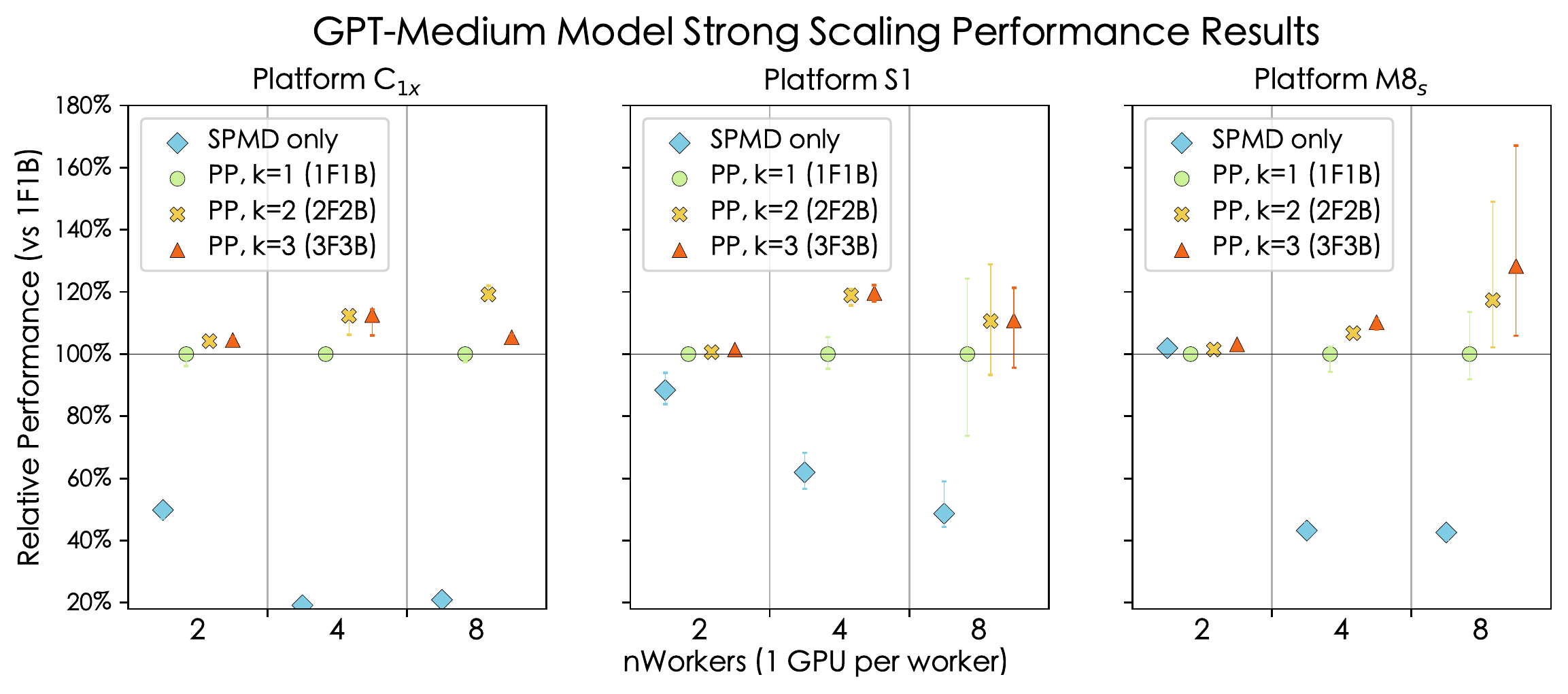}
	\caption{GPT-Medium strong scaling performance results on three target platforms, with different pipeline scheduling methods. SPMD only parallel is also tested for comparison. The error bars show the performance varying range of different steps.\label{fig:gpt_strong}}
\end{figure*}

\begin{figure}[t]
	\centering
	\includegraphics[width=0.48\textwidth]{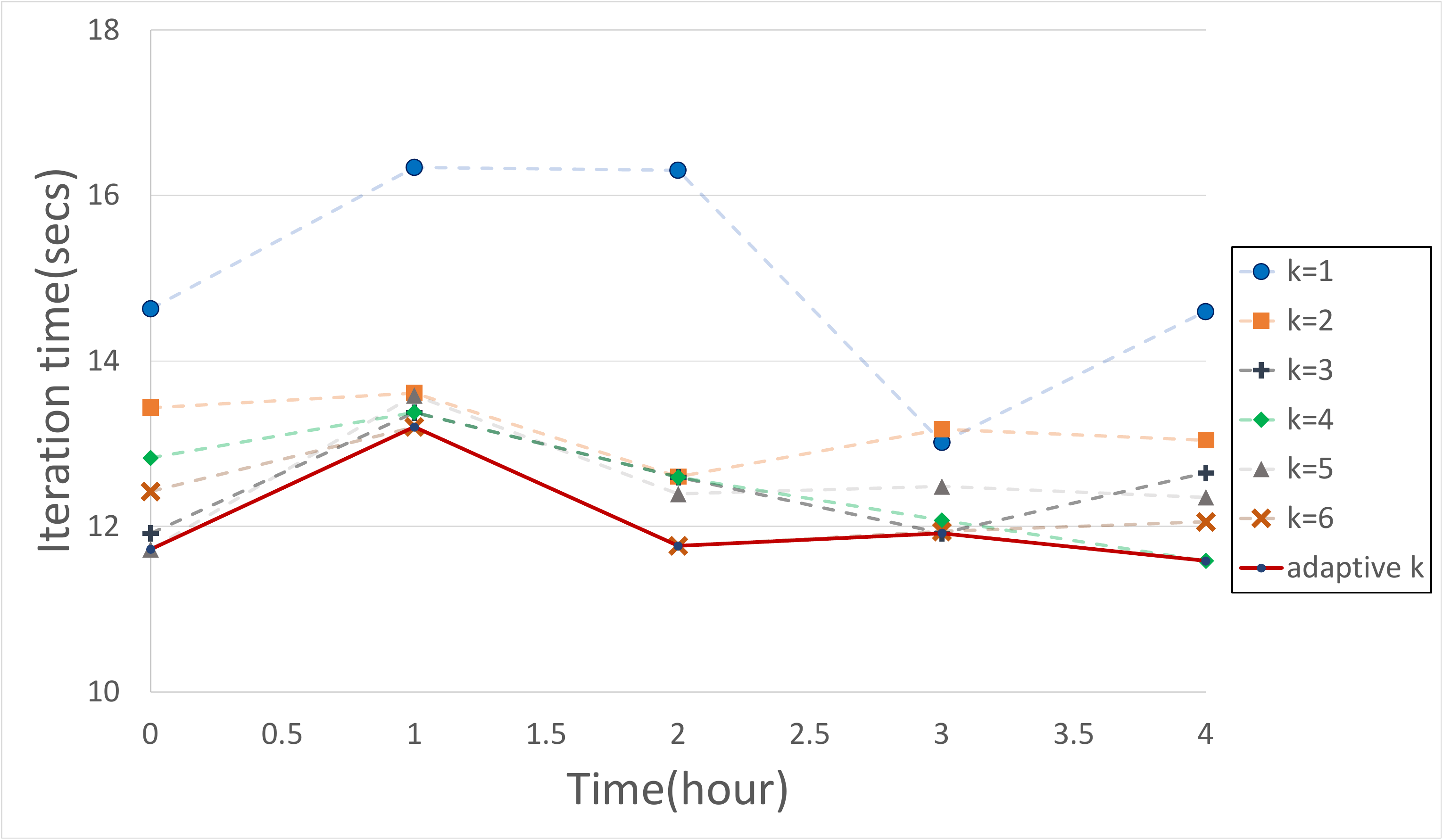}
	\caption{GPT adaptive tuning test. The tuning triggers once an hour and selects optimal plan among plans with different $k$ values, resulting in the tolerance in the ever-changing network. \label{fig:adaptive_tuning}}
\end{figure}

\subsection{Results}

\subsubsection{Pipeline Granularity Tests}

Pipeline granularity tests are examined on GPT-Medium model configuration, with pipeline parallel on 8 workers of Platform S1. It would be considered that k times of micro batches should be activated during a training iteration step of \OurApproach's \OurSchedule group schedule plan. Thus these granularity tests are designed for those memory limited models to verify that finer grained micro batches splitting could be used for \OurSchedule to preserve the maximum memory usage.

The tests set global batch size as fixed number 192, change k from 1 to 6, and changed the micro batch size as 6//k. Tests were arranged as 5 rounds of different time period in order to get more coverage under various level of overall cluster network load. The relative performance (compared to 1F1B of Round 1) and the variation records are shown in Fig. \ref{fig:gpt_granularity}.

The results could obviously indicate that \OurApproach got 10\%-25\% performance increase than 1F1B plan. It could also be speculated that network is busy during Round 3 and Round 5 test periods, because 1F1B timings fell down to ~90\% of Round 1. However the performance results of $k>1$ cases still kept at a much stable level and got more than 20\% improvement than 1F1B result, although they also slightly dropped. When $k \ge 3$, there is no deterministic further improvement while k becomes larger. That might reason on 1) no more overlapping under such higher pipeline granularity, 2) calculation of smaller micro batch would cause lower computing efficiency.

As the conclusion of tests above, \OurSchedule scheduling may make the chance of better much more stable computation and better communication overlapping which leads to parallel efficiency with finer granularity, while not bringing more memory needs.

\subsubsection{Weak Scaling Tests}

Weak scaling tests were done on U-Net and GPT models, up to 8 pipeline workers. For U-Net, global data batch size are set scaled with number of workers. Two configuration were tested individually on Platform M8$_s$, and the results in Fig. \ref{fig:unet_weak} shows the \OurApproach relative performance compared with 1F1B. UNet-Medium didn't have $k=4$ or N$_{workers}$=8 results because of OOM. More tensor communication could be found among the divided pipeline stages on U-Net structure, compared with layer based LM models like Bert or GPT. Despite this, up to 2-14\% performance improvement on UNet-Base and 4-5\% on UNet-Medium could still be observed when $k \ge 2$. 

For GPT, model parameter number scaling was performed for more workers as explained in Table \ref{tab:gpt_conf}. The global batch size is set as 64 for all tests. We also calculated out the achieved real FLOPs during the tests based on the method in \cite{megatron}. Tests were done on all three platforms to get better understand of the influence from different platform environment condition. The results are shown in Fig. \ref{fig:gpt_weak_tests}.

First, \OurApproach's \OurSchedule scheduling plan always outperformed 1F1B within the GPT weak scaling tests, especially in 8 worker cases. Focusing on this scale of three different platforms, it is significant that the performance on Platform C$_{1x}$ did not scale, mainly caused by it's narrow bandwidth getting opposite to higher communication demand for 8 stages. It is supposed that sharing jobs during training time on Platform M8${_s}$ might slightly influence the real communication latency and bandwidth. Thus FLOPs obtained on Platform S1 looks a little higher than M8$_s$, although they are connect with in the same production networking environment. This may also be used as one of reasons for why larger performance fluctuating could also be seen from the weak scaling results when number of workers increasing, which is same with all other tests.

\subsubsection{Strong Scaling Tests}

Strong scaling tests on GPT-Medium model could not only explain the \OurSchedule performance, we also tried to give a comparison with same scale SPMD only parallel training under same workload and platform conditions. These tests were also done on all three platforms, results shown in Fig. \ref{fig:gpt_strong}. The global batch size is set as 64 for all tests. The micro batch size is 1 for pipeline parallel tests and 8 for SPMD only tests. 

Similar with tests introduced in last sections, \OurSchedule showed up to 20\% performance improvement than 1F1B. Besides this, the results also indicates that for this GPT-Medium model configuration, pipeline almost always got higher performance on these platforms for production workloads.

For all SPMD parallel results, we checked the parallel strategies deduced by \PreviousWork automatic parallel system. The SPMD strategies are very data-parallel like, which needs about 0.7-1.4GB size data transferring during one micro batch calculation and is in line with our cognition. But this number is 2-5 times folded in pipeline parallel. So it is easy to understand why pipeline parallel should be a better parallel method under this case.

\subsubsection{Adaptive Tuning Test}

Fig~.\ref{fig:adaptive_tuning}  illustrates the tuning results of Ada-Grouper on Platform S1 whose network resource is preempted. Six different scheduling plans with k values ranging from 1 to 6 were generated by the Ada-Grouper. The online tuning triggers were set to occur once an hour. We sampled four hours in our experiment. The global batch size was kept constant at 192. By profiling the cross-stage communication and stage execution time, the performance of each plan is estimated. The points on the dotted lines indicate the estimated performance of individual plans at the evaluation time, while the point on the active line represents the optimal performance among all plans at the same evaluation time. The plan with $k=1$ is the original 1F1B plan which performs worse than almost all the other plans at each evaluation time due to the increased number of bubbles in the unstable network. While the other plans with different $k$ values from 2 to 6 display varied performances, the improvement does not necessarily become more prominent as the $k$ value increases. This is due to the requirement of a smaller micro-batch size as the $k$ increases, resulting in the under-utilization of computation efficiency and being constrained by the memory limit. Thus, the trade-off between computation efficiency and the capability of communication overlapping in preempted network leads to the necessity of estimating the overall performance.

In our experiment, \OurApproach initially chose $k=5$, followed by $k=6$ at the subsequent two evaluation times. Although the estimated performances between plans with $k=5$ and $k=6$ are close, they both performed significantly better than the 1F1B plan over 20\%. Network preemption is indicated to have been alleviated at the third hour, as 1F1B shows improvement on its performance compared to the past two hours, and the performances of all the other plans were very close. \OurApproach switched to the plan with $k=3$ at this time. At the last hour, the network became unstable again, leading to a larger performance gap between the different plans. The plan with $k=4$ achieves the optimal among all plans which surpasses 1F1B about 21\% , demonstrating that \OurApproach has the capability to deliver stable and satisfactory performances in an environment where preemption is present. The trigger frequency can be adjusted by users through environment variables, allowing for adaptation to frequently changing network environments.

\section{Related Works}

\subsection{Pipeline Parallelism}
GPipe \cite{gpipe} proposes training large deep models with pipeline parallelism and points out the peak memory usage and bubble problem. It uses gradient-checkpointing\cite{feng2021optimal} \cite{chen2016training} to alleviate the memory pressure. PipeDream \cite{pipedream} first proposes the 1F1B schedule plan and applies this technique to asynchronous training to reduce bubble time. Pipemare\cite{yang2021pipemare} provides a robust asynchorounous training for pipeline parallelism, which maintains model quality without sacrificing memory usage and utilization. DAPPLE \cite{dapple} extends 1F1B schedule plan on synchronous training and indicates the peak memory usage can stay at constant. Pipe-Torch\cite{zhan2019pipe} improves pipeline parallelism to hybrid parallelism method by combining data parallelism and model parallelism to accelerate training on heterogeneous network environment. Chimera\cite{li2021chimera} adopts the bidirectional pipeline parallelism to further improve pipeline efficiency. OOO\cite{oh2022out} proposes out-of-order scheduling for back-propagation, which facilitates distributed training. TeraPipe\cite{li2021terapipe} propose a novel pipeline approach from partitioning on token level dimensions for language model. 

\subsection{Deep Learning Compilers}
HLO\cite{xla} is an single assignment-based intermediate representation(IR) for tensor computations in XLA. MLIR~\cite{lattner2020mlir} is a reusable and extensible compiler infrastructure that standardizes the static single Assignment-based IR data structures and provides a declarative system to define IR dialects. Gshard\cite{gshard} and GSPMD\cite{gspmd} introduce collective communication primitives on HLO IR and provides convenient APIs for sharding large models. Relay \cite{roesch2019relay} presents a compiler framework to unify and generalize IR in existing frameworks. TVM \cite{chen2018tvm} is a compiler that exposes graph-level and operator-level optimizations to provide performance portability for DL workloads across diverse hardware backends.

\subsection{Communication Hiding Strategy}
PyTorchDDP\cite{li2020pytorch} implement data parallelism on PyTorch, which overlaps gradients reduction during back-propagation.
Horovod\cite{horovod} enables all-reduce gradients to be efficiently exchanged across multiple devices by employing a tensor fusion technique which optimizes both communication hiding and network utilization. HiDup\cite{zhang2022accelerating} duplicates the DNN model into two copies that have no dependency, and interleave their execution such that computation of one copy overlaps with communication of the other. IPart\cite{wang2021overlapping} overlaps gradient communication with backward computation and parameter communication by partitioning communication and computation in various partition sizes.

\subsection{Parallelism Framework}
Alpa\cite{zheng2022alpa}formulates a hierarchical model for parallelization, which models intra-operator parallelism as an ILP problem and develops a dynamic programming approach for exploring parallelism for both intra- and inter-parallelism. Unity\cite{unger2022unity} extends TASO\cite{jia2019taso} and introduces a parallel computing graph and applies randomized graph substitution to optimize algebraic transformations and parallelization simultaneously. FlexFlow\cite{lu2017flexflow} defines a "SOAP" search space and investigates parallelization by utilizing randomized search. Whale\cite{whale} partitions models according to the computation-balanced principle to adapt to heterogenous compute devices, allowing users to specify parallelization strategies with the help of parallelization primitives. Galvatron\cite{miao2022galvatron} proposes automatic parallel optimization for transformer-like models. Rhino\cite{zhang2023auto} is our previous work which automatically explores SPMD and pipeline parallelism by utilizing ILP and dynamic programming combined with data-driven heuristics, offering users flexible trade-offs between the search time and strategy quality. BytePS\cite{jiang2020unified} leverages spare CPUs and bandwidth resources in GPU clusters to accelerate distributed training with parameter server. Megatron\cite{megatron} supports training Transformer models at large scale with expert-designed parallelization strategies that combines data parallelism, tensor sharding parallelism and pipeline parallelism. DeepSpeed\cite{rasley2020deepspeed} is a high performance framework with expert-crafted strategies involving data and model paralellism. 
\section{Conclusion}

We present \OurApproach, an adaptive \OurSchedule scheduler that periodically adjusts the number of scheduling group members $k$ to accommodate to preempted network environment for pipeline parallelism. Firstly, we analyze the impact of the performance of 1F1B scheduling plan in a congested network, which is a common occurrence in cloud services. In comparison, the \OurSchedule plan results in fewer stalls for stage computation and a shorter pipeline length due to fewer bubbles.. Then, we analysis the capability to maintain a more stable and better performance in an unstable network environment. Additionally, the \OurApproach pass is introduced to generate all potential candidates with different group member count ($k$) and micro-batch size ($b$) given a fixed global batch size. To prune the candidate sets, we design an efficient enumeration algorithm to filter out all solutions that exceed memory limit or under-utilize device memory, resulting in potential candidates only exists on Pareto optimal frontier. Finally, to adapt to the ever-changing network environment, \OurApproach provides an automatic tuning mechanism to switch plans among candidates by estimating the performance of each plan. Our experiments demonstrate that \OurApproach achieves better performance on GPT and UNet on different platforms ranging up from 4\% to 30\% compared with 1F1B, while has the capability to maintain stable performance in the ever-changing network environment.

\bibliography{main}
\bibliographystyle{plain}

\end{document}